\documentclass{article}

\usepackage[activate={true,nocompatibility},final,tracking=true,kerning=true,spacing=true,factor=1100,stretch=10,shrink=10]{microtype}

\usepackage{arxiv}

\usepackage[font=small,labelfont={bf}, figurename=Figure, tablename=Table]{caption}

\usepackage[hidelinks]{hyperref}
\hypersetup{
  colorlinks   = true, 
  urlcolor     = esalightblue, 
  linkcolor    = esablue, 
  citecolor   = esablue 
}
\usepackage{amsmath}
\usepackage{amssymb}
\usepackage[utf8]{inputenc} 
\usepackage[T1]{fontenc}    
\usepackage{hyperref}       
\usepackage{url}            
\usepackage{booktabs}       
\usepackage{amsfonts}       
\usepackage{nicefrac}       
\usepackage{lipsum}
\usepackage{graphicx}
\graphicspath{ {./images/} }
\usepackage[nameinlink, capitalise]{cleveref}
\usepackage{xcolor}
\definecolor{esablue}{RGB}{0,57,158}
\definecolor{esalightblue}{RGB}{0,155,219}
\definecolor{esared}{RGB}{150,1,54}

\usepackage[giveninits=true,sorting=none,citestyle=numeric-comp]{biblatex}
\begin{document}

\title{Totimorphic structures for space application}

\author{
 Amy Thomas,\ \ Jai Grover,\ \ Dario Izzo,\ \ Dominik Dold \\[2pt]
  Advanced Concepts Team\\[1pt]
  European Space Agency, European Space Research and Technology Centre\\[1pt]
  Keplerlaan 1, 2201 AZ Noordwijk, The Netherlands\\
}

\maketitle
\begin{abstract}
We propose to use a recently introduced Totimorphic metamaterial for constructing morphable space structures. As a first step to investigate the feasibility of this concept, we present a method for morphing such structures autonomously between different shapes using physically plausible actuations, guaranteeing that the material traverses through valid configurations only while morphing. 
With this work, we aim to lay a foundation for exploring a promising and novel class of multi-functional, reconfigurable space structures.
\end{abstract}


\section{Introduction}
The last decade has seen rapid expansion and interest in the field of advanced materials and structures, especially in the development of programmable, multi-functional, and morphable structures. Such structures are particularly suited for space environments, where payload mass and volume are tightly constrained, making light structures capable of performing multiple functions highly desirable. For instance, Origami principles have already been used in the development of deployable solar panels and antennae, and applications for other advanced structures, such as NASA’s starshade \cite{starshade}, have been proposed \cite{ynchausti2022hexagonal,biswas2022ultra}. However, most deployable space structures are currently limited between strict configurations (typically stowed and deployed), often prohibiting any reconfiguration thereafter.

Here, we explore a recently proposed metamaterial called a \emph{Totimorphic structure} \cite{chaudhary2021totimorphic} whose characteristic properties might enable designs capable of redeployment and reconfiguration into many different shapes after initial deployment. This is especially intriguing for constructing adaptive structures, as many recent papers \cite{yazdani2022bioinspired,dold2023differentiable} have demonstrated the feasibility of changing a structure’s mechanical properties through geometric alterations alone. Such systems might enable mission designs capable of complex and efficient post-launch reconfiguration, adjusting to changing mission goals or conditions in situ and providing space missions with greater flexibility, thus fundamentally changing the types of missions possible in the future.

In the following, we first explain the concept of totimorphic structures before introducing a computational method for obtaining the actuations needed to morph them into different shapes.

\section{Methodology}
\begin{figure*}[t!]
    \centering
    \includegraphics[width=\textwidth]{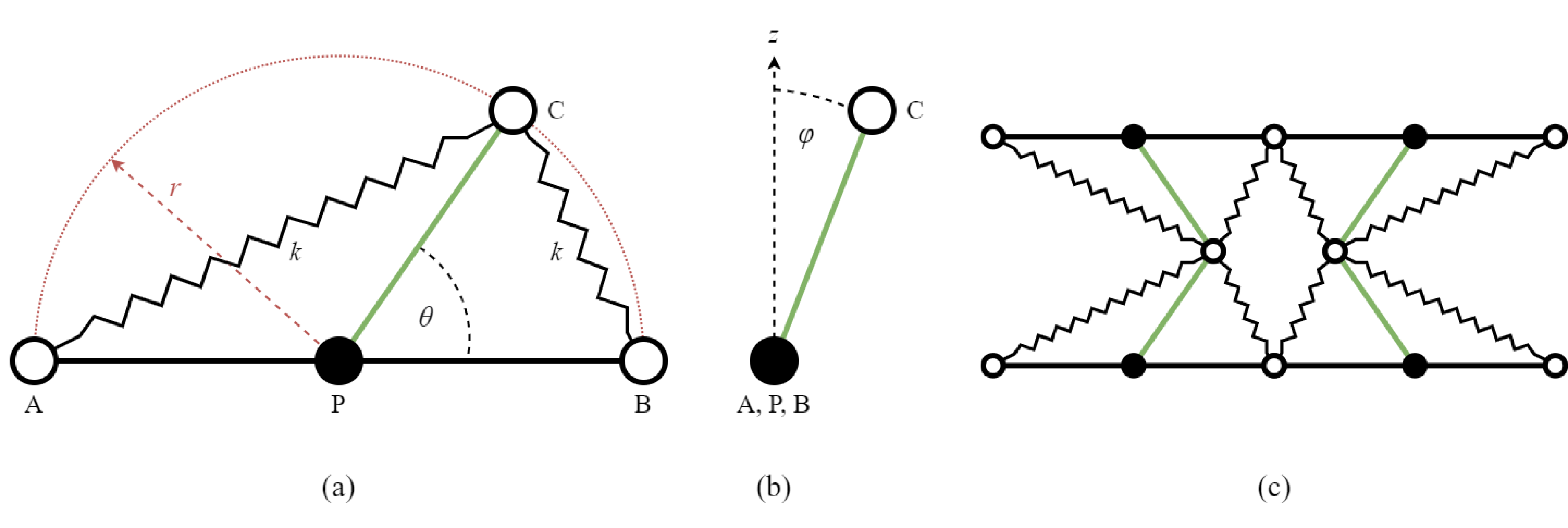}
	\caption{ \textbf{(a)} Illustration of a TUC with lever rotation $\theta$. \textbf{(b)} The TUCs beam rotation is given by $\varphi$, i.e., $A-P-B$ is seen from the side here. \textbf{(c)} A simple auxetic structure built from four TUCs.
	}\vspace{-4mm}	
	\label{fig:totimorphic}
\end{figure*}

Totimorphic structures are composed of neutrally-stable unit cells (also called Totimorphic Unit Cells, TUCs) [4] shown in \cref{fig:totimorphic}a,b. A TUC consists of a beam with a ball joint in its middle ($A-P-B$ in \cref{fig:totimorphic}a), a lever connecting to the joint ($P-C$) and two springs connecting the ends of the beam with the end of the lever ($A-C$ and $B-C$). The neutrally stable behaviour of TUCs arises from the lever-spring motive: if the two springs are zero-length springs with identical spring stiffness, any position of the lever results in zero moment acting on the lever. The result of this property is propagated across larger structures built from TUCs (e.g. \cref{fig:totimorphic}c), such that in the absence of external forces (e.g. gravity), the totimorphic structure will retain its shape while remaining completely compliant to any external force or displacement. By selective locking and unlocking of the structure’s joints, we predict that the totimorphic property will allow the structure to be smoothly morphed between different shapes via beam/lever rotations alone – which we exploit in our method presented in the next section. Once the desired target shape is reached, the structure can be made rigid by locking all joints.

Before presenting our approach for morphing such structures, we first introduce the used mathematical description. Each TUC is geometrically fully defined by its position vector $P$, the beam vector $\overrightarrow{AB}$, the lever length $r$, the angle between beam and lever $\theta$, and the roll angle of the lever from the vertical $\varphi$ (\cref{fig:totimorphic}b). TUCs may be connected to each other at the $A$, $B$, $C$ nodes, however for this paper’s analysis, only $A-B$ and $C-C$ connections were considered. Additionally, $r$ was set to be equal for all TUCs. If the coordinates of $A$, $B$ or $C$ are already known (i.e. the unit cell is attached to another unit cell), the unit cell can be described just by $\overrightarrow{AB}$, $\theta$, and $\varphi$ – which is useful for implementing the morphing method.

\section{Results}

Our morphing method works as follows. We first define the initial and target geometries, set a maximally allowed change in $\overrightarrow{AB}$, $\theta$, and $\varphi$ per iteration (same for all TUCs), and a subset of nodes are set as ‘pinned’ (i.e. their position is fixed, but they are allowed to rotate). We also ‘activate’ the pinned cells, (i.e. allow them to morph). Then we run the following iteration until the structure’s geometry is within some tolerance of the desired target geometry: each activated TUC changes $\overrightarrow{AB}$, $\theta$ and $\varphi$ as much as possible to reach its target configuration while still maintaining structural cohesion (i.e. no beams breaking/prolonging/squeezing or TUCs separating). Then they become ‘fixed’ (cannot be activated anymore in this iteration, so the TUCs are frozen in place) and inform their neighbouring cells that they should move next (they activate their neighbours). This process is repeated until all cells are fixed. Consequently, all TUCs are unfixed and the next iteration begins. Intuitively, a wave of ‘cell activation’ moves through the structure, where activated cells are moved closer to their target and become immovable (fixed) afterwards.
\begin{figure*}[t!]
    \centering
    \includegraphics[width=\textwidth]{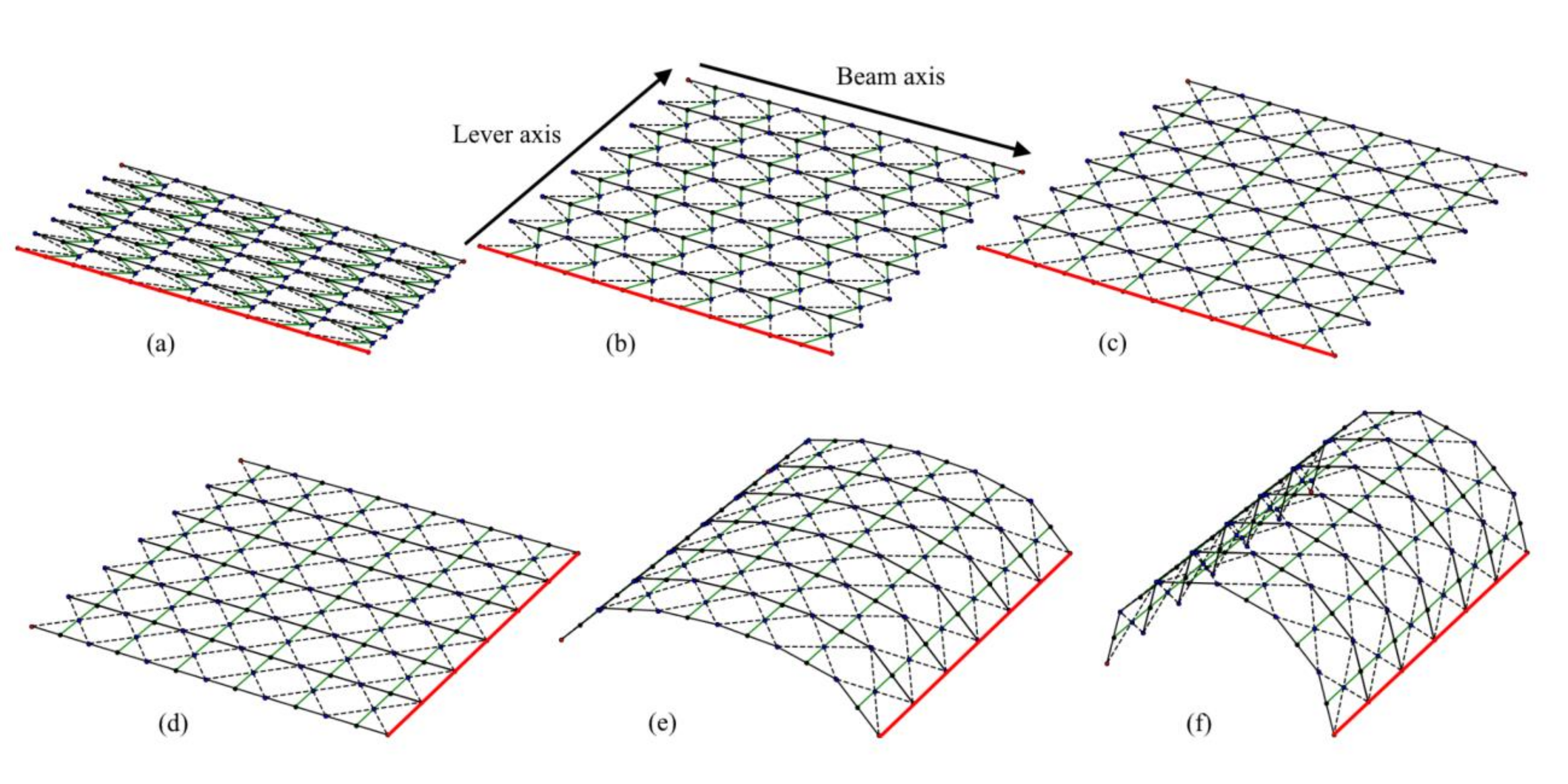}
	\caption{A totimorphic structure morphing continuously from a stowed plane to a square plane \textbf{(a)} $\theta=30$ \textbf{(b)} $\theta=60$ \textbf{(c)} $\theta=90$, and then into a semi-cylinder \textbf{(d-f)}. Red lines indicate pinned nodes with fixed coordinates.
	}\vspace{-4mm}	
	\label{fig:morphing}
\end{figure*}

The morphing method is illustrated in \cref{fig:morphing} for continuously morphing a structure to different target shapes: (a-c) turning a stowed geometry into a large flat surface, and (d-f) turning the flat surface into a half-cylinder. Although we halt at the shape in \cref{fig:morphing}f here, in principle the structure can be mapped to any developable surface of an equal or smaller surface area \cite{chaudhary2021totimorphic}, so long as there is a valid target geometry for the method.

In comparison with an Origami structure, which must necessarily unfold and change its thickness during deployment, we can see in (a-c) that the stowed totimorphic structure can continuously increase its surface area without any increase in the stowing thickness, resulting in a large stowed-to-deployed surface area ratio. Further, the stowing of the totimorphic structure does not constrain its final structural geometry, since it is possible for the structure to morph into intermediary states that enable better deployment before morphing into the final configuration.

\section{Discussion}

For the presented method to work, one requires knowledge about the exact initial and target configuration of the totimorphic structure (including angles) instead of just the shapes. Moreover, the choice of pinned nodes is crucial to ensure convergence to the target shape. Thus, we plan to develop an improved non-rigid morphing methodology that is both more robust and general, based on a model that predicts the whole structure’s response to small perturbations. Such a model might further allow us to use inverse design approaches from artificial intelligence \cite{dold2023differentiable} to autonomously find configurations that possess desired effective mechanical properties.

Although totimorphic structures are more flexible in their morphing capabilities than Origami-based structures, they come with the drawback of having a high number of degrees of freedom as well as many movable mechanical parts. We anticipate that this will lead to challenges in the production of a physical prototype, which have to be overcome first for totimorphic structures to compete with alternative approaches. For instance, the neutrally stable behaviour will most likely not be perfectly realised in a physical system due to effects such as bending of the beam in a TUC from radial forces applied through the lever – an effect that has to be considered during the design and testing stages of a real prototype.

We are confident that totimorphic structures are ideally suited for deployment in extremely-low gravity environments such as orbits and deep space, where no external loads due to gravity interfere with the morphing process and the neutrally stable behaviour can be utilised to its fullest. As noted in \cite{schenk2014zero}, the gravitational forces experienced on planetary bodies will impede the morphing of totimorphic structures, the extent of which has to be investigated in future work.

To lock and unlock the joints as well as induce beam/lever rotations, we envision a thin and foldable support layer glued to the totimorphic structure through which lock, unlock and rotation commands, e.g., from a microcontroller, can be relayed electronically. For instance, since only a small number of cells are unlocked at a time, electrical pulses could be used with a shared bus to realise an efficient and economic solution.

\section{Conclusion}

Totimorphic structures offer unique structural characteristics that make them ideal for space missions – providing a high degree of flexibility coupled with a low mass and volume. They are suitable for a multitude of potential space applications; such as deployable habitats and structures, tools (e.g. nets for grabbing something), or for building moving structures by creating locomotion from morphing. Hence, we believe that totimorphic structures are a promising candidate in the search for technologies that enable morphable, multi-functional space structures.

\section*{Acknowledgments}

We would like to thank Derek Aranguren van Egmond and Michael Mallon for helpful discussions, and our colleagues at ESA’s Advanced Concepts Team for their ongoing support. AT and DD acknowledge support through ESA’s young graduate trainee and fellowship programs. 

\printbibliography
\addcontentsline{toc}{section}{References}

\end{document}